\documentclass[aps,prl,reprint,showpacs,superscriptaddress]{revtex4-1}
\usepackage{amsmath,amssymb,graphicx}
\begin{document}
\title{Asymptotically exact probability distribution for the Sinai model with finite drift}

\author{Gareth Woods}
\affiliation{School of Physics and Astronomy, University of
Birmingham, Birmingham B15 2TT, United Kingdom}

\author{Igor V.\ Yurkevich}
\affiliation{School of Physics and Astronomy, University of
Birmingham, Birmingham B15 2TT, United Kingdom}

\author{Igor V.\ Lerner}
\affiliation{School of Physics and Astronomy, University of
Birmingham, Birmingham B15 2TT, United Kingdom}

\author{H. A. Kovtun}
\affiliation{B.\ Verkin Institute  for  Low  Temperature  Physics and Engineering, NASU,
Kharkiv 61103, Ukraine}

\date{\today}
\begin{abstract}
We obtain the exact asymptotic result for the disorder-averaged probability distribution function for a random walk in a biased Sinai model and show that it is characterized by a creeping  behavior of the displacement moments with time, $\left\langle x^n \right\rangle\sim t^{\mu n}$, where $\mu<1$ is dimensionless mean drift. We employ a method originated in quantum diffusion which is based on the exact mapping of the problem to an imaginary-time Schr\"{odinger} equation.
For nonzero drift such an equation has an isolated lowest eigenvalue separated by a gap from quasi-continuous excited states, and the eigenstate corresponding to the former governs the long-time asymptotic behavior.
\end{abstract}

\maketitle

The Sinai model  \cite{Sinai} is  one of the  simplest models for transport in  classical disordered systems  where physical behavior of various disorder-averaged quantities is interesting and non-trivial  and yet amenable to analytical calculations. The model describes random walks  under the influence of a spatially homogeneous thermal white noise in the presence of spatial disorder represented by a Gaussian quenched random-drift field $v({x})$. In one dimension (1D) such a  disorder strongly suppresses diffusion so that the mean random-walk displacement in the absence of the mean  drift  is given  by $\langle{x^2(t)}\rangle\propto \ln^4t $  \cite{Sinai,ClDifReview,*CDif}. In higher dimensions, where the properly generalized Sinai model describes various physical applications like hopping transport in the presence of charged or magnetic impurities, dynamics of dislocations in glasses, $1/f$-noise etc.,  diffusion can be either suppressed (sub-diffusion) for potential random drifts or enhanced (super-diffusion) for solenoidal ones \cite*{ClDifReview,*CDif,ArNel,*Dan2,KLY,*KLY:86c}. The model  also  exhibits mesoscopic fluctuations  \cite{KLY} within the ensemble of realizations  similar to the mesoscopic fluctuations  \cite{BLA:85,*L+S:85} in quantum diffusion in the Anderson model. In such a situation, any exact analytical result for the Sinai model can be potentially of a wide interest and applicability. Naturally, the 1D case is most promising for finding exact solutions.

The exact analytic  results for the most generic quantity in the 1D Sinai model, the random walks probability distribution function (PDF),  has been presented by Kesten  \cite{Kesten} but only in the limit of zero mean drift  while the extension of this method to the case of an arbitrary  drift is not known. Although some exact results  for the 1D Sinai model with an arbitrary drift have been obtained --  e.g.\  return probability  \cite{ClDifReview,*CDif}, mean first-passage time \cite{GB:98,*LDMF:08,*LDC:01}, or persistence   \cite{Comtet:02} -- none of the methods used for these quantities has been generalized for the PDF.

In the present publication we close this gap by applying the method developed \cite{FPY,*FPY1,*YL:99b} in the context of quantum diffusion in the presence of losses (non-Hermiticity) to obtain exact asymptotic (long-time) results for the PDF of the 1D Sinai model in the presence of a finite drift.

The essence of the method is the following. First we obtain a formally exact solution for the Laplace image of the PDF, $\mathcal{P}_\varepsilon ({x,x'})$, for a given realization of the quenched random drift field $v({x})$. We represent this solution in terms of two linearly  independent functions, $\xi_\varepsilon ^\pm$, which obey certain boundary conditions only on the left or only on the right boundary of the 1D system, respectively. This allows us to formulate the Langevin-type equations for $\xi_\varepsilon ^\pm$, with $x$ serving as an \textit{effective time variable} (and the Laplace variable $\varepsilon $ being a parameter).  Then we use the Furutsu--Novikov  \cite{Furutsu,*Novikov:65} technique for averaging over the quenched disorder, expressing the result in terms of the eigenvalues of an auxiliary Fokker--Planck equations (FPE). Finally, we map FPE to an equivalent imaginary-time Schr\"{odinger} equation (SE) and obtain an asymptotically exact expression  (in the long-time limit, i.e.\ for $\varepsilon \to0$) for the disorder-averaged PDF, $P_\varepsilon (x-x')$.

The peculiarity of the solution for nonzero drift is that the SE is characterized by a two-scale potential and thus has an isolated lowest eigenvalue separated by a gap from quasi-continuous excited states. The PDF for the biased Sinai model is mostly contributed by the former, with the latter giving the main contribution to the PDF for zero drift. This explains inevitable difficulties for extending standard methods successful for the unbiased Sinai model to the model with nonzero drift.

The 1D Sinai model (in dimensionless variables  \cite{WYKL1}) is described by the following Langevin equation:
\begin{align}\label{LE}
    \frac{{\mathrm{d} x} }{{\mathrm{d}} t}= \mu+v{(x)}+\eta({t})\,.
\end{align}
Here $\eta({t})$ is the Gaussian thermal noise with $\overline{\eta({t})}=0$ and  $\overline{\eta({t}) \eta({t'})}=2 \delta({t-t'})$ (\,$\overline{\phantom{,}\dots\phantom{i}}$ denotes the thermal-noise averaging),  $\mu$ is a mean drift velocity due to, e.g., a constant drag force and $v(x)$ is a quenched Gaussian random-drift field  with
\begin{align}\label{vv}
    \langle v(x) \rangle &=0\,,& \langle v(x)\,v(x') \rangle =2\delta({x-x'})\,.
\end{align}

After averaging over the thermal noise,  the Sinai model is described in terms of the Fokker-Planck equation:
\begin{align}\label{FP}
\bigr({\partial_t -\partial_x {\mathcal{D}}_x }\bigl){\cal P}(x,x';t)&=\delta({x-x'})\delta({t})\,,
\end{align}
Here ${\mathcal{D}}_x\equiv\partial_x-\mu-v({x})\,,$ and  $\mathcal{P}(x,x';t)$  is the PDF in a given realization of the quenched random field $v({x})$, obeying the  boundary  conditions $ {\mathcal{D}}_x{\cal P}(x,x';t)\bigr|_{x=\ell _{\pm}}=0 $ at the sample boundaries $\ell _\pm\to\pm\infty$.

Introducing the Laplace transform of the PDF,
\begin{align}
{\cal P}_{\varepsilon}(x,x')=\int_0^{\infty}{\rm d}t\,\mathrm{e}^{-\varepsilon t}{\cal P}(x,x';t)\,,
\end{align}
we rewrite Eq.~(\ref{FP}) as follows,
\begin{align}\notag
\bigl(\varepsilon-\partial_x{\mathcal{D}}_x\bigr)
{\cal P}_{\varepsilon}(x,x')=\delta(x-x')\,,
\end{align}
which can be formally solved by the substitution
\begin{align}\notag
&{\cal P}_{\varepsilon}(x,x')=\theta(x\!-\!x') {\cal P}^+_{\varepsilon}(x,x')
+\theta(x'\!-\!x) {\cal P}^-_{\varepsilon}(x,x')\,,\\
&\;\label{Psigma}\\[-12pt]
&{\cal P}^{\sigma}_{\varepsilon}(x,x')=\frac{\varphi _\varepsilon^\sigma(x)\varphi _\varepsilon^{-\sigma}(x')}
{\varphi_\varepsilon ^{+}(x')\partial_{x'}\varphi_\varepsilon ^{-} ({x'})-\varphi_\varepsilon ^{-}(x')\partial_{x'}\varphi_\varepsilon ^{+} ({x'})}\notag\,.
\end{align}
Here  $\varphi_{\varepsilon}^\sigma(x)$ (with $\sigma=\pm$) are the eigenfunctions of the homogeneous equation which satisfy the boundary conditions only on the left ($\varphi _\varepsilon^-$) or only on the right ($\varphi _\varepsilon^+$):
\begin{align}\notag
\bigl(\varepsilon-\partial_x{\mathcal{D}}_x\bigr)
\varphi_{\varepsilon}^\sigma(x)&=0\,;& {\mathcal{D}}_x\varphi _\varepsilon^\sigma(x)\bigr|_{x=\ell_{\sigma}}&=0\,.
\end{align}

Our main task is to find the full PDF, ${P}(x,x';t)=\langle \mathcal{P}(x,x';t) \rangle $,  by performing the ensemble averaging over the Gaussian random-drift field  (\ref{vv}). To this end, we introduce  the shifted logarithmic derivatives of the eigenfunctions $\varphi _\varepsilon ^\sigma$:
\begin{align}\label{xi}
\sigma\xi^{\sigma}_\varepsilon(x)\equiv
{\partial_x\left[\ln\varphi _\varepsilon^\sigma(x)\right]} -\mu-v(x)\,.
\end{align}
They obey the following Riccati equations
\begin{align}\label{Ric}
\partial_x\xi^{\sigma}_\varepsilon(x)=\sigma\left[ (\xi^{\sigma}_\varepsilon)^2
-\varepsilon\right]- \bigl[\mu+v(x)\bigr]\,\xi^{\sigma}_\varepsilon(x)
\,,
\end{align}
with one-sided boundary condition for each of them, $\xi^{\sigma}_\varepsilon(x=\ell_{\sigma})=0$. Then $x$ plays a role of an effective time variable and Eqs.~(\ref{Ric}) a role of Langevin equations describing the `time' evolution of $\xi ^-(x)$ and the time `anti-evolution' for $\xi ^+({x})$ (as the boundary condition fixes the latter at a maximal value of  $x$). For real and positive $\varepsilon$ the functions $\xi ^\sigma$ are positive everywhere as $\xi ^-$ increases from $0$ and $\xi ^+$ decreases to $0$,  as follows from  Eq.~(\ref{Ric}).

\begin{figure}    \includegraphics[width=.9\columnwidth]  {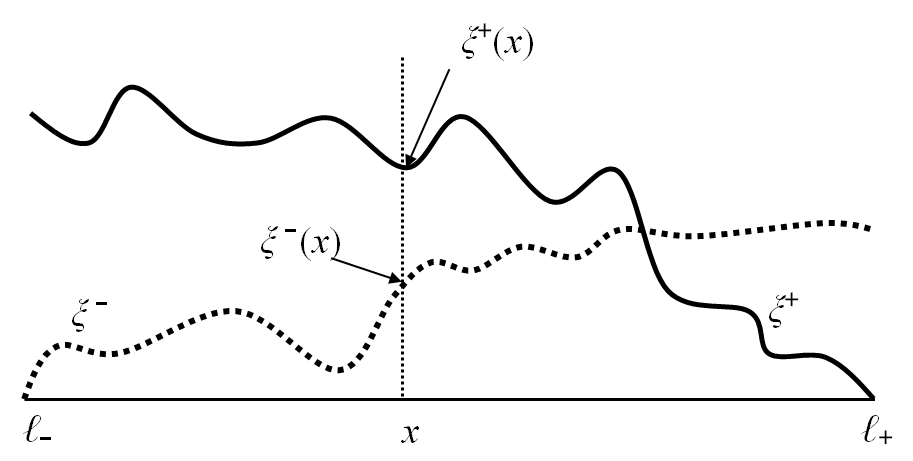}  \caption{The stochastic variable $\xi^-(x')$ depends on realization of random drifts for the `earlier time' ($<x$) while $\xi^+(x)$ for the `later' ($>x$). Therefore, blocks built of $\xi^+$ and $\xi^-$ are statistically independent.}\label{Xi}\end{figure}

Now we can represent the PDF in Eq.~(\ref{Psigma}) in terms of $\xi_\varepsilon ^\sigma$.  First, it follows from the definition (\ref{xi}) that
\begin{align}\label{Ps}
    P_\varepsilon^\sigma(x,x') = \frac{\exp\Bigl\{\int_{x'}^x \textrm{d}y\big[\xi^+(y)+v(y)+\mu\big]\Bigr\}
    }{\xi^{+}_\varepsilon(x')+\xi_{\varepsilon }^-(x')}\,.
    \end{align}
Secondly, we  get rid of the explicit dependence on $v(x)$ in Eq.~(\ref{Ps}) by using Eq.~(\ref{Ric}) where we divide both parts by $\xi ^\sigma$ and integrate over $x$ which allows us to connect the integrals of $\xi ^\sigma$ and over $1/\xi ^\sigma$ resulting in
\begin{align}
{\cal P}^{\sigma}_{\varepsilon}(x,x')=\frac{\xi^{\sigma}_\varepsilon(x')\, /\,\xi^{\sigma}_\varepsilon(x)}
{\xi^{+}_\varepsilon(x')+\xi_{\varepsilon }^-(x')}\,\exp\biggl[-\sigma\varepsilon\!\int\limits_{x'}^{x}\frac{{\rm d}x_1}{\xi^{\sigma}_\varepsilon(x_1)}\biggr]\,.
\end{align}
Finally, we exponentiate the denominator of the pre-exponential factor in Eq.~(\ref{Ps}) as $\int\nolimits_{0}^{\infty}{\mathrm{e}^{-y(\xi ^++\xi ^-)}}\mathrm{d}y $ which allows us to represent $\mathcal{P}^{\sigma}$ as the product of $\xi ^+$ and $\xi ^-$ functions. Since the (anti)causality of the Riccati equations (\ref{Ric}),  $\xi ^-(x')$ depends, in any realization of the random drift field $v(x)$ only  on $v({x})$ with the argument   $x<x'$ while $\xi ^+(x')$ only on those which the argument $x>x'$, as illustrated in Fig.~\ref{Xi}.
This allows us to  average over the Gaussian quenched disorder, Eq.~(\ref{vv}), separately for $\xi ^\pm$ functions -- in essence, this is the well-known Furutsu -- Novikov \cite{Furutsu,*Novikov:65} technique. Thus we  represent the averaged $P^\pm$ functions as follows ({with $\sigma=\pm$}):
\begin{align}\label{Pav}
P^{\sigma}_{\varepsilon}(x,x')=-\frac{\sigma}{\varepsilon}\, \partial_x\int\limits_{0}^{\infty}{\rm d}y\,R^{\sigma}(y;x,x')\,Q ^{-\sigma}(y;x')\,,
\end{align}
where
\begin{align}\notag
R^{\pm}(y;x,x')&=-\partial_y\left\langle \exp\biggl[-y\xi^{\pm}_\varepsilon(x')\mp\varepsilon \!\! \int\limits_{x'}^{x}\frac{{\rm d}x_1}{\xi^{\pm}_\varepsilon(x_1)}\biggr]\right\rangle\,,\\[-2pt]\,\label{RQ} \\[-6pt]\notag
Q^{\pm}(y;x')&=\left\langle \mathrm{e}^{ -y\,\xi^{\pm}_\varepsilon(x')}\right\rangle.
\end{align}
The fully averaged PDF, $P_\varepsilon ({x,x'})$, is expressed in terms of  $P_\varepsilon ^\pm ({x,x'})$, in the same way as the PDF for a given realization, ${\mathcal{P}}_\varepsilon ({x,x'}) $, in terms of   ${\mathcal{P}}_\varepsilon ^\pm ({x,x'})$,   Eq.~(\ref{Psigma}).

 \begin{subequations}\label{QR1}
 The causality of the Riccati equations allows us  to derive in a standard way  \cite{LGP} the  Fokker-Planck type equations for the partial distribution functions $Q^{\sigma} $ and $R^\sigma$:
 \begin{align}\label{Q}
\sigma \partial_{x'} Q^{\sigma}&=\left( \varepsilon y-{\hat{M}^{\sigma} \partial _y }\right) Q^{\sigma},\\\notag\hat{M}^{\sigma}& \equiv y(y+1)\partial _y+y(1+\sigma\mu); \\
\sigma \partial_{x'} R^{\sigma}&=\left({   \varepsilon y-\partial _y\hat{M}^{\sigma}}\right) R^{\sigma} \equiv \hat{L}^{\sigma}R^{\sigma} \,.\label{R}
\end{align}
\end{subequations}
Equation (\ref{Q})  is solved with the `initial' condition for $x'$,
$
Q ^{\sigma}(y;x'\!=\!\ell_{\sigma})=1,
$
and the  boundary conditions for the auxiliary variable $y$,
$
Q^{\sigma}(y\!=\!0;\,x')=1 $ and \mbox{$ Q^{\sigma}(y\!\to\!\infty;\,x')=0
$}.
The appropriate standard solution for  $x'$  far from the physical boundaries, i.e.\ $|x'-\ell _\sigma|\gg1$, goes over to the stationary, $x'$-independent function:
\begin{align}\label{Qsol}
Q^{\sigma}(y;x')\to Q^{\sigma}(y)=\frac{{\rm K}_{\mu}\! \left[ \, 2\sqrt{\varepsilon(1+y)}\, \right]}{{\rm K}_{\mu}(2\sqrt{\varepsilon})}(1+y)^{\frac{\sigma\mu}2},
\end{align}
where $K_\mu({z})$ is the modified Bessel function.

The operator ${\hat L}^{\sigma}$ in  Eq.~(\ref{R}) is Hermitian with respect to the scalar product defined with the weight function
\begin{align}\label{wf}
    w (y)=(1+y)^{\sigma\mu+1}
\end{align}
and it has only positive eigenvalues.  The solution to Eq.~(\ref{R}) must be everywhere finite, integrable,  monotonic with $y$ and satisfy the `initial' condition
$R^{\sigma}(y;x'\!=\!0)=-\partial_yQ^{\sigma}(y)$, as
 follows from the definition (\ref{RQ}).  Thus it can be formally represented as $
    R^{\sigma}(y;x,x')=\mathrm{e}^{-(x-x')\sigma\,{\hat L}^{\sigma}}\,\partial_yQ^{\sigma}(y;x')\,$.
Far from the physical boundaries, i.e.\ at $|x'\!-\!\ell _\sigma|\gg1$, the solution for $Q^{\sigma}(y;x')$ is $x'$-independent, Eq.~(\ref{Qsol}), so that $R^{\sigma}$ becomes translationally invariant, $R^{\sigma}(y;x,x') \to R^{\sigma}(y;x-x')$. Then we choose $x'=0$ in Eqs.~(\ref{Pav}) and  (\ref{Psigma}) thus obtaining   $P^{\sigma}_{\varepsilon}(x)\equiv P^{\sigma}_{\varepsilon}(x,x'=0)$ as follows:
\begin{align}\label{Psig}
P^{\sigma}_{\varepsilon}(x)=\frac{\sigma}{\varepsilon}\, \partial_x\int\limits_{0}^{\infty}{\rm d}y\,Q^{-\sigma}(y) \,\mathrm{e}^{-x\sigma\,{\hat L}^{\sigma}}\,\partial_yQ^{\sigma}(y)
\end{align}

The translational invariance of $R^\sigma$ ensures a useful relationship, ${\hat L}^{\sigma}R^{\sigma}_{\varepsilon}(y,x=0) =\sigma\,\varepsilon\,Q^{\sigma}_{\varepsilon}(y)$, which follows from substituting $\partial_{x'}=-\partial_{x} $ into Eq.~(\ref{R}) and putting $x=0$ after applying $\partial_{x}$ to the definition (\ref{RQ}).
 Substituting this into
Eq.~(\ref{Psig})  allows us to reproduce immediately a well-known result  \cite{ClDifReview} for the Laplace transform of the return probability ${\cal R}_{\varepsilon}\equiv P^{\sigma}_{\varepsilon}(0)$:
$$
{\cal R}_{\varepsilon}=\int\limits_{0}^{\infty}{\rm d}y\,Q^{-\sigma}Q^{\sigma}=\frac{{\rm K}_{\mu+1}(2\sqrt{\varepsilon}){\rm K}_{\mu-1}(2\sqrt{\varepsilon})}{{\rm K}^2_{\mu}(2\sqrt{\varepsilon})}-1.
$$
For an arbitrary $x$, it is convenient to represent the integral in Eq.~(\ref{Psig}) in terms of the eigenvalues of the operator $\hat{L}^{\sigma}$, defined by  the eigenvalue equation
\begin{align}\label{EVE}
    \hat{L}^{\sigma}R^{\sigma}_E(y)=ER^{\sigma}_E(y)\,.
\end{align}
First we solve Eq.~(\ref{EVE}), considering $E$ as a given external parameter,  in the two limits,  $y\gg 1$ and $y\ll\varepsilon^{-1}$. We are interested in the long-time asymptotic behavior corresponding to the Laplace variable $\varepsilon \ll1$. Therefore, these two limiting solutions should match each other in the asymptotically wide region $1\ll y\ll\varepsilon^{-1}$. This matching  allows us to find the actual eigenvalues $E({\varepsilon })$.

The two limiting solutions for an arbitrary $E$ are
 \begin{subequations}\label{2limits}
\begin{align}
R^{\sigma}_E(y\ll\varepsilon^{-1})&=\,_{2}{\rm F}_1(\kappa+r,\kappa-r;1;-y)\,,\label{a} \\
R^{\sigma}_E(y\gg 1)&=y^{-\kappa}{\rm K}_{2r}(2\sqrt{\varepsilon y})\,,\label{b}\\\label{r}
\kappa=1+\tfrac{1}{2}\sigma\mu&\,, \quad r =\sqrt{\tfrac{1}{4}\mu^2-E}\,,
\end{align}
where $_{2}{\rm F}_1$ is the hypergeometric function.
\end{subequations}
To match these solutions in the region $1\ll y\ll\varepsilon^{-1}$, we expand Eq.~(\ref{a}) for $y\gg1$ up to the leading and the first sub-leading term, which yields
\begin{subequations}\label{1-2}
\begin{align}\label{1}
R^{\sigma}_E\approx  \frac{\Gamma(2r) y^{-\kappa+r}}{\Gamma(\kappa+r)\Gamma(1-\kappa-r)}  + (r\to-r) \,,\end{align}
and similarly Eq.~(\ref{b}) for $\varepsilon y\ll1$, which yields
\begin{align}\label{2}
R^{\sigma}_E\approx     \frac{\pi }{2\sin2\pi r}\left[\frac{(\varepsilon y)^{-\kappa-r}}{\Gamma(1-2r)}-\frac{(\varepsilon y)^{-\kappa+r}}{\Gamma(1+2r)}\right].
\end{align}
\end{subequations}
The terms $y^{-\kappa-r}$ and $(\varepsilon y)^{-\kappa+r}$ are the main sub-leading terms in Eqs.~(\ref{1}) and (\ref{2}), respectively, provided that  $\operatorname{Re}r\!<\!1/2$ --   which we will show to hold at $0\!<\!\mu\!<\!1$  \cite{WYKL2}.
In this case, matching the coefficients in Eqs.~(\ref{1}) and (\ref{2}) gives the following eigenvalue equation:
\begin{align}\label{spectrum}
\varepsilon^{2r}=\frac{{\rm\Gamma}^2(2r)\,{\rm\Gamma} (-r-\sigma\mu/2)\,{\rm\Gamma}(1-r+\sigma\mu/2)} {{\rm\Gamma}^2(-2r)\,{\rm\Gamma}(r-\sigma\mu/2)\, {\rm\Gamma}(1+r+\sigma\mu/2)}\,.
\end{align}
There is a quasi-continuum set of eigenvalues, corresponding to the imaginary $r$, which is given for $|r|\ll \mu\ne0$ (when the r.h.s.\ of Eq.~(\ref{spectrum}) is approximately $1$) by
\begin{align}\label{Em}
E_n^{\sigma}=\frac{\mu^2}{4}+\left(\frac{\pi nm}{\ln\varepsilon}\right)^2\,, \quad n=1,2 \dots
\end{align}
This expression is valid only for $m\ll|\ln \varepsilon |$ but the lower part of the spectrum is all we need to determine the long-distance behavior.  Note that for $\mu=0$ and $|r|\ll1$, the r.h.s.\  of Eq.~(\ref{spectrum}) equals $-1$, which leads to the shift $n\to n+1/2$ in the spectrum (\ref{Em}). Substituting   the asymptotic eigenfunctions (\ref{2limits}) with these values of $E$ into Eq.~(\ref{Psig}) recovers the known result for $\mu=0$  \cite{Kesten}:
$$
P_{\varepsilon}(x)=\frac{4}{\pi\varepsilon\, \ln^2(\varepsilon)}\sum_{n=0}^{\infty}\frac{(-1)^n}{n+\frac{1}{2} } \exp\biggl[-\bigg(\frac{\pi\big(n+\frac{1}{2} \big)}{\ln{\varepsilon}}\bigg)^{\!2} |x|\biggr].
$$
However, the main difference between the biased and unbiased case is the appearance of an additional bound eigenstate of ${\hat L}^{+}$ with $E\ll\mu^2/4$, corresponding to a real positive value of the parameter $r$, Eq.~(\ref{r}), close to $\mu/2$. In this case, the l.h.s.\ of Eq.~(\ref{spectrum}) goes to $0$, while $\Gamma(r-\mu/2)$ in the denominator of the r.h.s.\ is close to the pole. Expanding it around the pole and substituting $r=\mu/2 $ elsewhere, we find the lowest eigenvalue:
\begin{align}\label{mu1}
E_0(\varepsilon)=  {\pi\mu}\big[{\sin\pi\mu}\,{\rm\Gamma}(\mu)\big]^{-2}\varepsilon^{\mu} \,.
\end{align}

To illustrate finding the eigenstates, it is useful to map the eigenvalue equation (\ref{EVE}) for $\hat{L}^+$ into an equivalent Schr\"{odinger} equation with the imaginary `time' $x$,
\begin{align}\label{SE}
    -\partial_x\Psi (x,\,z)=\left[-\partial^2_z+U (z)\right] \Psi (x,\,z) \,,
\end{align}
with the help of the substitution
\begin{align}\notag
    y\to y(z) &\equiv\sinh^2\frac{z}{2}\,, & R^{+}&\to \frac{\Psi (x,\,z)}{y(z)^{\frac{1}{4}}[1+y(z)]^{\frac{3+2 \mu}{4}} }\,.
\end{align}
\begin{figure}    \includegraphics[width=.9\columnwidth]  {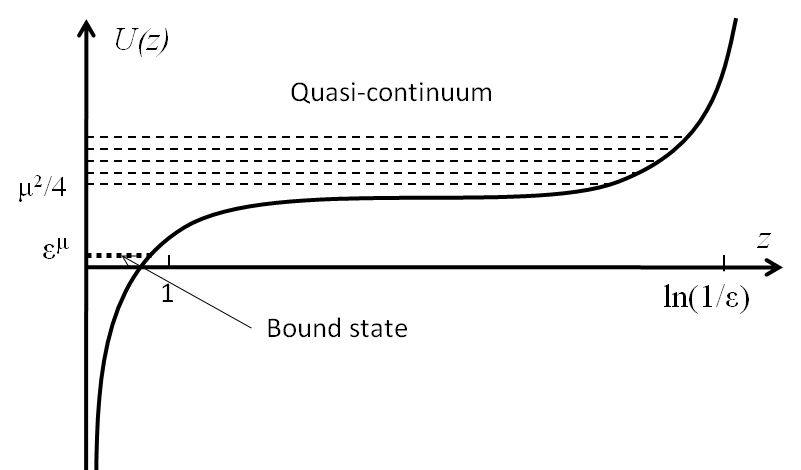}  \caption{The effective potential for the  Schr\"{odinger} equation.}\label{SEq}\end{figure}
The effective potential in Eq.~(\ref{SE}) is given by
\begin{align*}
    U (z)=\frac{\mu^2}{4}-\frac{1}{16y(z)}- \frac{(1+2\mu)(3+2\mu)}{16[1+y(z)]}+\varepsilon y
\end{align*}
and is illustrated in Fig.~\ref{SEq}. It underlines the main difference between the biased ({$\mu\ne0$ }) and unbiased ({$\mu=0$}) cases: in the former case the effective potential is characterized by two scales which results in the emergence of the bound state separated by a gap of order $\mu^2/4$ from the quasi-continuum.

As a result, only the lowest eigenvalue of Eq.~(\ref{mu1}) contributes to the large-scale behavior (at $x\gg 4/\mu^2$) of  $P^{+}_{\varepsilon}(x)$ while $P^-$ is of no interest: the distribution function in this limit is one-sided. Thus in the general expansion for $\partial_yQ^{+}(y;x)$,
\begin{align}\notag
\partial_yQ^{+}(y;x)=c_0\,R_0^+(y)\,\mathrm{e}^{-E_0x} +\sum_{n}\,c_n\,R_{n}^{+}(y)\,\mathrm{e}^{-\sigma E^{+}_nx}
\end{align}
we can keep only the corresponding eigenstate $R_0^+$ obtained by substituting $E\to E_0$ in Eq.~(\ref{1-2}). The coefficients in the above expansion are found as the appropriate scalar products with the weight functions (\ref{wf}).  In the asymptotic limit of Eq.~(\ref{1-2}) we find $c_0=-\mu^{-1}$. This leads, with $E_0({\varepsilon })$ given for $0\!<\!\mu\!<\!1$ by Eq.~(\ref{mu1}), to
\begin{align}\label{final}
P^{+}_{\varepsilon}(x)=\varepsilon^{-1}\,E_0(\varepsilon) \,\mathrm{e}^{-E_0(\varepsilon)\,x}\,.
\end{align}
This is the main result of this paper.  As an example it
gives asymptotically exact expressions for the moments:
\begin{align}\label{moments}
\langle x^n(t)\rangle = \frac{{\rm \Gamma}(n+1)}{{\rm \Gamma}(\mu n+1)}\,\left[\frac{\pi\mu{\rm \Gamma}^2(\mu)}{\sin\pi\mu}\right]^n \,t^{\mu n}.
\end{align}
This describes a  `creeping' behavior but the propagation is considerably faster than in the unbiased case (where, e.g., $\langle{x^2(t)}\rangle\propto \ln^4t$).  The results of Eqs.~(\ref{final}) and (\ref{moments}) have already been known  \cite{ClDifReview,BerSch:86} for a related model described by the master equation on a 1D lattice, $\dot P_n=W_nP_{n-1}-W_{n+1}P_n$, with a broad distribution of the hopping probabilities $W$, which diverges at small $W$ as $\psi({W})\sim W^{\mu-1}$. It has been conjectured, albeit with some misgivings  \cite{ClDifReview}, that these results might be applicable to the biased Sinai model. Here we have proved that Eq.~(\ref{final}) is asymptotically exact.

Let us finally note   Eq.(\ref{final}) is valid, with an appropriate expression for $E_0(\varepsilon )$, for any $\mu$. Thus for $1\!<\!\mu\!<\!2$,  the expansion of $E_0$ in $\varepsilon$
starts from a linear term,
\begin{align}\label{final1}
E_0(\varepsilon)=\frac{\varepsilon}{\mu-1}+b_{\mu}\,\varepsilon^{\mu}.
\end{align}
This means that in this case  there is a constant-velocity drift, while the correction term defines the anomalous sub-diffusive dispersion but  precise form of $b_\mu$ is  of a relatively little importance.

\begin{acknowledgments}
This work has been supported by the EPSRC grant  T23725/01. HAK is thankful for  hospitality extended to her in Birmingham at the  initial stage of this work.
\end{acknowledgments}


\begin{thebibliography}{23}
\bibitem [{WYK(){\natexlab{a}}}]{WYKL1}%
  \BibitemOpen
  \bibinfo {note} {In proper dimensional
  variables, the thermal noise correlator is proportional to the bare
  (short-range) diffusion coefficient, while the random-drift correlator to
  some constant characterizing the disorder strength}\BibitemShut {NoStop}%
\bibitem [{WYK(){\natexlab{b}}}]{WYKL2}%
  \BibitemOpen
   \bibinfo {note} {When $\operatorname{Re}r
  >1/2$, one would need to keep $y^{-\kappa+r-1}$ instead of $y^{-\kappa-r}$ in
  Eq.~(\ref{1}) which is, however, beyond the accuracy of the derivation of
  Eqs.~(\ref{2limits}).}\BibitemShut {Stop}%
\end{thebibliography}

%

\end{document}